\documentclass[11pt,a4paper]{article}

\usepackage[truedimen,margin=32mm]{geometry} 

\usepackage{mathrsfs}
\usepackage{amssymb}
\usepackage{amsmath}
\usepackage{ascmac}
\usepackage{amsthm}
\usepackage[pdftex]{graphicx}
\usepackage[pdftex]{color}
\usepackage{natbib}
\usepackage{color}
\usepackage{setspace}
\usepackage{url}

\usepackage{titlesec}
\titleformat*{\section}{\large\bfseries}
\titleformat*{\subsection}{\it}

\newtheorem{algo}{Algorithm}

\def\ep{{\varepsilon}}

\def\th{{\theta}}

\def\thh{{\widehat{\th}}}

\def\beh{{\widehat{\beta}}}
\def\sih{{\widehat{\sigma}}}

\title{{\bf Adaptively Robust Geographically \\
Weighted Regression}}
\date{}

\begin{document}

\maketitle
\doublespacing

\vspace{-1.5cm}
\begin{center}
{\large Shonosuke Sugasawa$^{*}$\footnote{Corresponding author, Address: 5-1-5, Kashiwanoha, Kashiwa, Chiba 2778568, JAPAN, Email: sugasawa@csis.u-tokyo.ac.jp} and Daisuke Murakami$^{\dagger}$}
\end{center}

\medskip
\noindent
$^{*}$Center for Spatial Information Science, The University of Tokyo\\
$^{\dagger}$Department of Data Science, The Institute of Statistical Mathematics

\medskip
\medskip
\begin{center}
{\bf \large Abstract}
\end{center}

\vspace{-0cm}
We develop a new robust geographically weighted regression method in the presence of outliers.
We embed the standard geographically weighted regression in robust objective function based on $\gamma$-divergence.    
A novel feature of the proposed approach is that two tuning parameters that control robustness and spatial smoothness are automatically tuned in a data-dependent manner. 
Further, the proposed method can produce robust standard error estimates of the robust estimator and give us a reasonable quantity for local outlier detection.  
We demonstrate that the proposed method is superior to the existing robust version of geographically weighted regression through simulation and data analysis.


\newpage
\section{Introduction}
In the era of open data, traffic volume, crime counts, land cover, and many other spatial data. These data often have outliers. For instance, traffic counts can be extremely large due to congestion after a car accident, while crime counts can be huge in specific districts. Measurement errors and typing errors can produce outliers as well. It is an emergent task to develop regression approaches that flexibly handle outliers in spatial data.
The objective of this study is to develop a novel approach to robustify geographically weighted regression (GWR; \cite{brunsdon1998geographically}) against outliers. GWR is a local regression approach for modeling spatially varying coefficients, which has been widely accepted in environmental science (e.g., \cite{lin2011using}), epidemiology (e.g., \cite{dong2018geographically}), and other fields. Yet, the basic GWR is highly sensitive to outlier \citep{GWRbook}. Thus, as illustrated in \cite{lesage2004family}, the local regression estimates can be “contaminated” by an outlier in the local samples.

\cite{GWRbook} suggested two approaches that robustify GWR. The first approach performs GWR after removing samples taking extremely large residuals in an initial fit. This practical approach is explored by \cite{harris2010robust} and extended by \cite{harris2014multivariate} for outlier detection. The second approach down-weights samples with large residuals through iterative GWR fitting. This down-weighting approach has widely been extended. For instance, \cite{lesage2004family} proposed a Bayesian GWR with non-constant variance that regularizes (or down-weights) outliers through priors. The Bayesian GWR has been applied to econometric analysis \citep{ma2020geographically}, regional development analysis \citep{clifton2006institutions}, and forest analysis \citep{subedi2018bayesian}. Other robust estimation, which assigns less weights for outliers, has also been applied. For example, \cite{zhang2011local} applied a least absolute deviation (LAD) estimation, \cite{salvati2012small} applied a M-quantile estimation, and \cite{chen2012geographically} applied an asymmetric absolute loss-based estimation for geographically weighted quantile regression.
There are other approaches to handling outliers under spatial dependence \citep[e.g.][]{mur2007outliers}, such methods cannot be directly applied to the GWR framework due to the spatial heterogeneity of regression coefficients in GWR.

The main drawbacks of above mentioned existing robust approaches that we focus on are mainly two points. 
First, existing methods only consider robust estimation of spatially varying parameters but do not care much about the robust selection of the bandwidth parameter crucially related to final results.
Secondly, although existing methods entail some tuning parameters that control robustness against outliers, the selection of these parameters is not carefully discussed; they are simply determined by rule-of-thumb.
Since we do not know how many outliers are included in practice, such tuning parameters should be tuned based on the observed data; otherwise, the estimation method can be statistically inefficient or may not be robust enough to produce stable estimates.  
To overcome the drawbacks, we introduce a new framework for robust estimation and inference of GWR. 
The primary tool for our proposal is $\gamma$-divergence \citep{jones2001comparison,fujisawa2008robust}, which is a general class of robust objective functions that contains the log-likelihood function as a particular case. We will demonstrate that the resulting estimating equations for regression coefficients can be seen as weighted estimating equations in which potential outliers are automatically down-weighted. 
In this sense, the proposed approach has some connection with the existing method by \cite{harris2010robust}.  
We will demonstrate that the robust objective function can be efficiently optimized via the Majorization-Minimization algorithm \citep[e.g.][]{hunter2004tutorial}. 
Owing to the statistically coherent framework, we can define a robust cross-validation method to select the optimal bandwidth from data. Furthermore, we can also derive a robust estimation of asymptotic standard errors of estimated coefficients. 
The $\gamma$-divergence includes a tuning parameter that controls the robustness of the objective function, and we propose a data-dependent selection strategy of the tuning parameter via recently proposed criterion \citep{sugasawa2021tuning} that can be easily implemented in practice. 
Therefore, in the proposed method, the amount of robustness and spatial bandwidth are automatically determined in a data-dependent manner. 
Such a feature is a clear advantage over the existing robust approach.

This paper is organized as follows.
In Section \ref{sec:method}, we introduce the proposed robust methods and provide estimation algorithms.
In Section \ref{sec:sim}, we evaluate the numerical performance of the proposed methods together with some existing methods through simulation studies.
In Section \ref{sec:app}, we demonstrate the proposed method through spatial regression modeling of crime data in Tokyo. 
Finally, we give some discussions in Section \ref{sec:disc}.

\section{Robust Geographically Weighted Regression}\label{sec:method}

\subsection{Settings and robust objective function}
Let $y_i$ be a response variable and $x_i$ is a vector of covariates in the $i$th location, for $i=1,\ldots,n$, where $n$ is the number of samples.
We suppose we are interested in estimating the spatially varying model: $y_i\sim N(x_i^\top\beta_i, \sigma_i^2)$, where $\beta_i$ and $\sigma_i^2$ are spatially varying regression coefficients and error variances. 
Further suppose that location information $s_i$ (e.g. longitude and latitude) is also available for the $i$th location.
These varying parameters can be estimated via geographically weighted likelihood given by 
\begin{equation}\label{LL}
(\beh_i,\sih_i^2)={\rm argmax}_{\beta,\sigma^2}\ L_i(\beta,\sigma^2), \ \ \ \ 
L_i(\beta,\sigma^2)=\sum_{j=1}^n w(\|s_i-s_j\|/b)\log \phi(y_j;x_j^\top\beta,\sigma^2),
\end{equation}
where $w(\cdot)$ is a weight function and $b$ is a bandwidth.
Given $\sigma^2$, the maximization of (\ref{LL}) with respect to $\beta$ is equivalent to minimizing $\sum_{j=1}^n w(\|s_i-s_j\|/b)(y_j-x_j^\top\beta)^2$, which is the standard GWR method.

A practical limitation of using the geographically weighted method (\ref{LL}) is the potential sensitivity against outliers. 
If some influential outlier exists at some location, the estimation accuracy of the parameters around the location may worsen.
To make the estimation procedure robust against outliers, we replace the log-likelihood function in (\ref{LL}) with the robust divergence. 
In particular, we adopt a transformed version of the $\gamma$-divergence \citep{fujisawa2008robust} given by 
\begin{align}
D_i(\beta,\sigma^2)&=\frac1{\gamma}\log\left(\sum_{j=1}^n w_{ij}(b) \phi(y_j;x_j^\top\beta,\sigma^2)^{\gamma}\right) \notag \\
& \ \ \ \ \ \ \ +\frac1{1+\gamma}\log\left\{\sum_{j=1}^n w_{ij}(b) \int \phi(t;x_j^\top\beta,\sigma^2)^{1+\gamma}dt\right\} \notag \\
&=\frac1{\gamma}\log\left(\sum_{j=1}^n w_{ij}(b) \phi(y_j;x_j^\top\beta,\sigma^2)^{\gamma}\right)  +\frac{\gamma}{2(1+\gamma)}\log\sigma^2 + C  \label{gam-div},
\end{align}
where $w_{ij}(b)=w(\|s_i-s_j\|/b)$ and $C$ is an irrelevant constant independent of the parameters. 
We note that 
$$
\lim_{\gamma\to 0}\left\{D_i(\beta,\sigma^2)-\gamma^{-1}\sum_{j=1}^n w_{ij}(b)\right\}
=L_i(\beta,\sigma^2),
$$
thereby the maximization of (\ref{gam-div}) with respect to $\beta$ and $\sigma^2$ is almost the same as that of (\ref{LL}) when $\gamma$ is close to $0$. 
Hence, (\ref{gam-div}) can be regarded as a natural extension of the standard geographically weighted method.
Here $\gamma$ is a tuning parameter controlling the robustness of the resulting estimator, and we provide a data-dependent selection method later.

Note that the first order condition of the minimizer of (\ref{gam-div}) with respect to $\beta$ is given by 
\begin{equation}\label{WEE}
\sum_{j=1}^nw_{ij}(b)\phi(y_j;x_j^\top\beta,\sigma^2)^{\gamma}x_j(y_j-x_j^\top \beta)=0.
\end{equation}
The above equation reduces to the estimating equation of the standard GWR when $\gamma=0$.
With non-zero $\gamma$, the equation (\ref{WEE}) is so called ``weighted estimating equation", and $\phi(y_j;x_j^\top\beta,\sigma^2)^{\gamma}$ is supposed to ``downweight" outliers in that its value will be small when $y_j$ is an outlier, namely, the absolute residual $|y_j-x_j^\top\beta|$ is relatively large.
In particular, when $|y_j-x_j^\top\beta|\to\infty$, it follows that $\phi(y_j;x_j^\top\beta,\sigma^2)^{\gamma}|y_j-x_j^\top\beta|\to 0$ as long as $\gamma>0$, so that the outlier information is automatically ignored in the equation (\ref{WEE}).

\subsection{Estimation algorithm}
Although the functional form of (\ref{gam-div}) seems complicated, we can efficiently minimize $-D_i(\beta,\sigma^2)$ using a Majorization-Minimization (MM) algorithm \citep[e.g.][]{hunter2004tutorial} similar to one given in \citep{kawashima2017robust}.
Using Jensen's inequality, the negative objective function can be evaluated as follows: 
\begin{equation}\label{MM}
-D_i(\beta,\sigma^2) \leq \frac1{2\sigma^2}\sum_{j=1}^n u_{ij}^{\dagger}(y_j-x_j^\top\beta)^2+\frac{1}{2(1+\gamma)}\log \sigma^2 + C^{\ast}, 
\end{equation}
where $C^{\ast}$ is a constant independent of $\beta$ and $\sigma^2$, $\beta_{\dagger}$ and $\sigma_{\dagger}^2$ are some fixed values of $\beta$ and $\sigma^2$, respectively.
Here $u_{ij}^{\dagger}$ is known as ``normalized weight" given by  
$$
u_{ij}^{\dagger}
=\frac{w_{ij}(b) \phi(y_j;x_j^\top\beta_{\dagger},\sigma_{\dagger}^2)^{\gamma}}
{\sum_{\ell=1}^n w_{i\ell}(b)\phi(y_\ell;x_\ell^\top\beta_{\dagger},\sigma_{\dagger}^2)^{\gamma}},
$$
which does not depend on the parameters, $\beta$ and $\sigma^2$, but on the fixed values, $\beta_{\dagger}$ and $\sigma_{\dagger}^2$.
Since the majorization function in (\ref{MM}) is similar to the log-likelihood function of the normal linear regression models, the updating steps for $\beta$ and $\sigma^2$ are analytically obtained. 
The MM algorithm iterates the computation of the majorization function and minimization of the majorization function, which is summarized as follows:

\begin{algo}[MM algorithm]\label{algo:MM}
For $i=1,\ldots,n$, starting with some initial values, $\beta_{i(0)}$ and $\sigma_{i(0)}^2$, repeat the following two steps until convergence: 
\begin{itemize}
\item[-]
(Computation of normalized weights) 
$$
u_{ij}^{(s)}
=\frac{w_{ij}(b) \phi(y_j;x_j^\top\beta_{i(s)},\sigma_{i(s)}^2)^{\gamma}}
{\sum_{\ell=1}^n w_{i\ell}(b)\phi(y_\ell;x_\ell^\top\beta_{i(s)},\sigma_{i(s)}^2)^{\gamma}}, \ \ \ \ j=1,\ldots,n
$$
\item[-]
(Update of $\beta$ and $\sigma^2$) 
\begin{align*}
&\beta_{i(s+1)}\  \leftarrow  \ \left(\sum_{j=1}^n u_{ij}^{(s)}x_jx_j^\top\right)^{-1}\sum_{j=1}^n u_{ij}^{(s)}x_jy_j\\
&\sigma_{i(s+1)}^2 \ \leftarrow \ (1+\gamma)\sum_{j=1}^n u_{ij}^{(s)}(y_j-x_j^\top\beta_{i(s+1)})^2
\end{align*}
\end{itemize}
\end{algo}

We may use the estimates from the standard GWR method as a reasonable initial value in Algorithm \ref{algo:MM}.
We note that the updating step for $\beta_i$ reduces to the standard GWR estimator under $\gamma=0$, since $u_{ij}^{(s)}|_{\gamma=0}=w_{ij}(b)$.

\subsection{Selection of the tuning parameters}
The proposed method has two tuning parameters, $b$ (bandwidth) and $\gamma$ (robustness control), which should be selected in a data-dependent manner.

First, we consider a selection strategy for $b$ with a fixed value of $\gamma$.
Note that the standard cross-validation technique used in GWR is not appropriate since the squared loss is sensitive to outliers. 
Instead, we use a robust criterion based on the $\gamma$-divergence. 
Let $\beh_{i(-i)}$ and $\sih_{i(-i)}^2$ are the estimators based on (\ref{gam-div}) without $(x_i,y_i)$. 
Then, we define a robust cross validation (RCV) criterion for $b$ as 
\begin{equation}\label{RCV}
{\rm RCV}(b; \gamma) = \frac1{\gamma}\log\left(\sum_{i=1}^n  \phi(y_i;x_i^\top\beh_{i(-i)},\sih_{i(-i)}^2)^{\gamma}\right)  +\frac{\gamma}{2(1+\gamma)}\log\left(\sum_{i=1}^n\sih_{i(-i)}^2\right),
\end{equation}
where the optimal $b$ is the maximizer of ${\rm RCV}(b)$. 
When $\gamma$ is close to $0$, the maximization of RCV is almost the same as maximizing $\sum_{i=1}^n \log \phi(y_i;x_i^\top\beh_{i(-i)},\sigma_{i(-i)}^2)$, which is equivalent to minimizing $\sum_{i=1}^n (y_i-x_i^\top\beh_{i(-i)})^2$ when $\sih_{i(-i)}^2$ is constant over $i$.
Hence, the criterion (\ref{RCV}) is a natural generalization of the standard cross-validation technique adopted in GWR.

We next consider selecting the robustness parameter $\gamma$.
We here suggest using an asymptotic approximation of the Hyvarinen score \citep{sugasawa2021tuning} with $\gamma$-divergence and the model assumption, $y_i\sim N(x_i^\top\beta_i, \sigma_i^2)$, defined as 
\begin{equation}\label{H-score}
H(\gamma; b)
=\sum_{i=1}^n\frac{1}{\sih_{i(\gamma)}^4}\bigg[2\left\{\gamma(y_i-x_i^\top\beh_{i(\gamma)})^2-\sih_{i(\gamma)}^2\right\}w_{i(\gamma)}
+(y_i-x_i^\top\beh_{i(\gamma)})^2w_{i(\gamma)}^2\bigg],   
\end{equation}
where $w_{i(\gamma)}=\phi(y_i;x_i^\top\beh_{i(\gamma)},\sih_{i(\gamma)}^2)^{\gamma}$, and $\beh_{i(\gamma)}$ and $\sih_{i(\gamma)}$ are estimates of $\beta_i$ and $\sigma_i^2$, respectively, under robustness parameter $\gamma$ and bandwidth parameter $b$.
It should be noted that most of the weights $\{w_{i1}(b),\ldots,w_{in}(b)\}$ can take almost $0$ when $b$ is small so that the actual sample size used in the estimation of $\beta_i$ and $\sigma_i^2$ can be small, which may break the justification using the criterion (\ref{H-score}).

To take into account the property of (\ref{H-score}), we propose the following simultaneous selection strategy for $\gamma$ and $b$.

\begin{algo}[Selection of $\gamma$ and $b$]
First prepare candidate sets, $\gamma\in \Gamma=\{\gamma_1,\ldots,\gamma_J\}$ and $b\in B=\{b_1,\ldots,b_L\}$, where $\gamma_1\leq \cdots \leq \gamma_J$ and $b_1\leq \cdots \leq b_L$.
The selection procedure consists of the following two steps: 
\begin{itemize}
\item[-]
(selection of $\gamma$) \ \ 
Using the largest bandwidth, namely, $b=b_L$, compute $\gamma_{opt}={\rm argmin}_{\gamma\in \Gamma}H(\gamma; b_L)$.

\item[-]
(selection of $b$) \ \
Compute $b_{opt}={\rm argmax}_{b\in B}{\rm RCV}(b; \gamma_{opt})$.
\end{itemize}
\end{algo}

\subsection{Standard error calculation}\label{sec:SE}
Let $\sum_{j=1}^n F_{ij}$ be the left-hand side of the estimating equation (\ref{WEE}). 
From the standard theory of estimating equations, the asymptotic covariance matrix of $\beh_{i(\gamma)}$ as the solution of the estimating equation (\ref{WEE}) can be estimated as 
$$
\widehat{{\rm Var}}(\beh_{i(\gamma)})=J_i(\beh_{i(\gamma)}, \sih_{i(\gamma)})^{-1}I_i(\beh_{i(\gamma)}, \sih_{i(\gamma)})J_i(\beh_{i(\gamma)}, \sih_{i(\gamma)})^{-1},
$$
where 
\begin{align*}
J_i(\beta, \sigma^2)
&\equiv 
\sum_{j=1}^{n}\frac{\partial F_{ij}}{\partial \beta^{\top}}
=\sum_{j=1}^n w_{ij}(b)\phi(y_j;x_j^{\top}\beta,\sigma^2)^{\gamma}\left\{\frac{\gamma(y_j-x_j^{\top}\beta)^2}{\sigma^2}-1\right\}x_jx_j^{\top}, \\
I_i(\beta, \sigma^2)
&\equiv \sum_{j=1}^n F_{ij}F_{ij}^{\top}
=\sum_{j=1}^n 
w_{ij}(b)^2\phi(y_j;x_j^\top\beta,\sigma^2)^{2\gamma}(y_j-x_j^\top \beta)^2x_jx_j^{\top}.
\end{align*}
Note that when $\gamma=0$, the above asymptotic covariance formula reduces to one for the standard GWR method.

\subsection{Outlier detection}
A desirable feature of the proposed method is to give a natural outlier detection strategy, using the density power term appeared in (\ref{WEE}).
Let $\beh_i$ and $\sih_i$ be the robust estimator with $b=b_{\rm opt}$ and $\gamma=\gamma_{\rm opt}$. 
We then define the normalized weight as 
\begin{equation}\label{weight}
U_i=\frac{\phi(y_i;x_i^\top\beh_i,\sih_i^2)^{\gamma_{\rm opt}}}{n^{-1}\sum_{j=1}^n\phi(y_j;x_j^\top\beh_j,\sih_j^2)^{\gamma_{\rm opt}}},
\end{equation}
where we note that $\sum_{i=1}^n U_i=n$.
If $y_i$ is an outlier in the sense that the density value $\phi(y_i;x_i^\top\beh_i,\sih_i^2)^{\gamma_{\rm opt}}$ is extremely small compared to the other observations, the weight $U_i$ is also small.  
Therefore, the small value of $U_i$ is a reasonable signal that $y_i$ is an outlier.

\section{Simulation Studies}\label{sec:sim}
 
We present simulation studies to illustrate the performance of the proposed method and existing robust and non-robust approaches through simulation studies. 
First, we uniformly generated $n=500$ spatial locations $s_1,\ldots,s_n$ in the domain $\{s=(s_1,s_2)\ | \ s_1\in [-1,1], \ s_2\in [0,2], \ s_1^2+0.5s_2^2>(0.5)^2\}$.
Then, we let  $z_1(s_i)$ and $z_2(s_i)$ be the two independent realizations of a spatial Gaussian process with mean zero and a covariance matrix defined from an isotropic exponential function: ${\rm Cov}(z_k(s_i), z_k(s_j))=\exp(-\|s_i-s_j\|/\phi)$, $k=1,2$, where $\phi$ is the range parameter. 
We considered two cases of the parameter, $\phi=0.4$ and $0.8$.
Then, we define two covariates $x_1(s_i)$ and $x_2(s_i)$ via linear transformations $x_1(s_i)=z_1(s_i)$ and $x_2(s_i)=rz_1(s_i)+\sqrt{1-r^2}z_2(s_i)$ with $r=0.75$, which allows dependence between $x_1(s_i)$ and $x_2(s_i)$.
Note that the value of spatial range $\phi$ is related to the strength of spatial correlation in the covariates, that is, a larger value of $\phi$ leads to covariates that are more likely to hold spatial collinearity.
To quantify the degree of collinearity, we use the condition number \citep[e.g.][]{brunsdon2012living} which computes the ratio of the largest eigenvalues to the smallest eigenvalues of $\sum_{j=1}^n w_{ij}x_jx_j^\top$ at the $i$th location, where $x_{j}=(x_1(s_j), x_2(s_j))^\top$ and $w_{ij}$ is the spatial weight between $i$th and $j$th locations.   
We here adopt the Gaussian kernel $w_{ij}=\exp(-\|s_i-s_j\|^2/2h^2)$ with bandwidth $h$.
Since it is suggested that the condition numbers above around 30 indicates considerable collinearity \citep[e.g.][]{belsley2005regression}, we calculated the number of locations whose condition number is greater than 30 with various values of $h$.
The results are shown in Table \ref{tab:NC}, form which we can observe that considerable spatial collinearity exist even under large values of bandwidth $h$ under $\phi=0.8$.

The response at each location is generated from the following model: 
$$
y(s_i) = \beta_0(s_i) + \beta_1(s_i)x_1(s_i) + \beta_2(s_i)x_2(s_i) +\ep(s_i), \ \ \ i=1,\ldots,n,
$$
where $\ep(s_i)$'s are error terms and are mutually independent.
We consider the following two scenarios of the distribution of $\ep(s_i)$:
\begin{align*}
&\text{Scenario (I)} \ \ \ep(s_i)\sim (1-\omega) N(0,\sigma^2) + \omega N(0, a^2\sigma^2),  \\
&\text{Scenario (II)} \ \ \ep(s_i)\sim (1-\omega) N(0,\sigma^2) + \omega N(a,\sigma^2), 
\end{align*}
where we set $\sigma^2=1$, $a=10$ and $\omega\in\{0, 0.05, 0.1, 0.15\}$.
Here $\sigma^2$ is the error variance of non-outlying observations, $\omega_i$ is the outlier ratio and the second component generates outliers. 
Outliers are distributed around $0$ in scenario (I), while outliers in scenario (II) are always positive, and the error distribution is not symmetric. 
Regarding the regression coefficients, we independently generated from a Gaussian spatial process. 
We set that all the processes have a zero mean and isotropic exponential function given by 
$$
{\rm Cov}(\beta_k(s_i), \beta_k(s_j))=\tau^2\exp\left(-\frac{\|s_i-s_j\|}{\psi_k}\right), \ \ \ \ k=0, 1, 2, 
$$
where $\psi_k$ is the range parameter, and $\tau^2$ is the variance parameter. 
We fix $\tau^2=2$ and $\psi_k=k+1$ in our study.

\begin{figure}[!htb]
\centering
\includegraphics[width=12cm,clip]{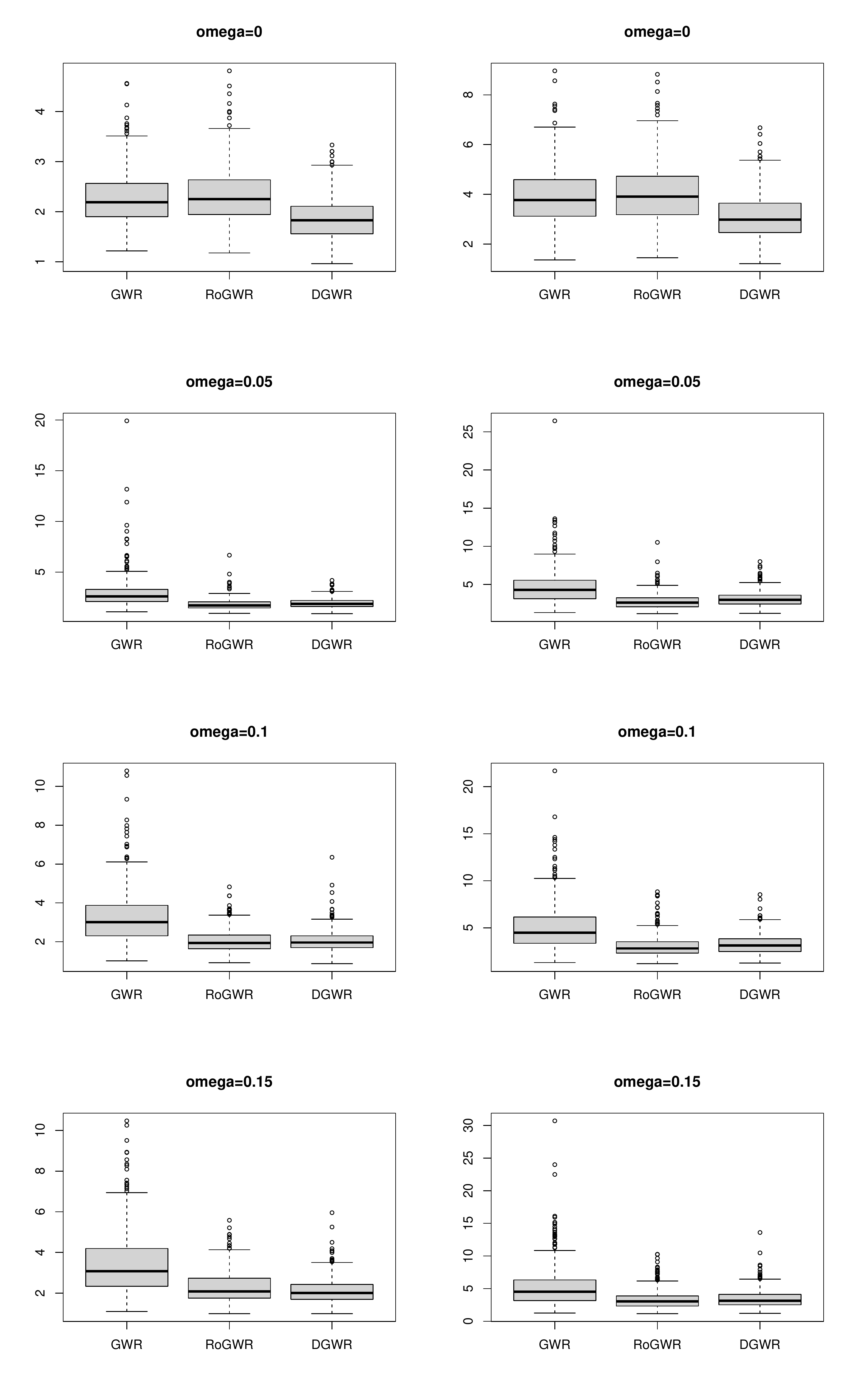}
\caption{Boxplot of MSEs for GWR, RoGWR and DGWR based on 500 simulated datasets under scenario (I) with two cases of spatial range parameter to generate covariates, $\phi=0.4$ (left) and $\phi=0.8$ (right).
\label{fig:sim1}
}
\end{figure}

\begin{figure}[!htb]
\centering
\includegraphics[width=12cm,clip]{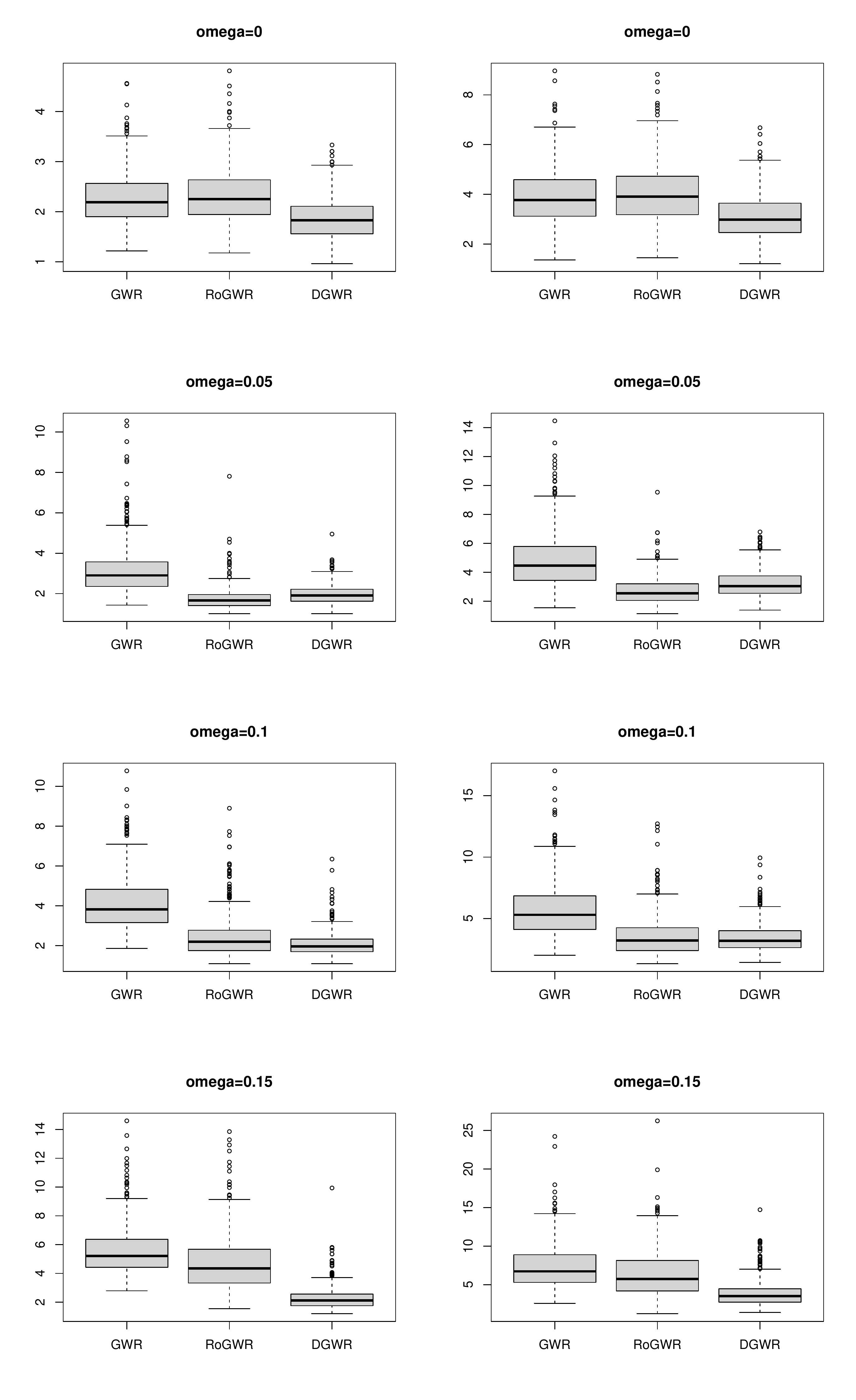}
\caption{Boxplot of MSEs for GWR, RoGWR and DGWR based on 500 simulated datasets under scenario (II) with two cases of spatial range parameter to generate covariates, $\phi=0.4$ (left) and $\phi=0.8$ (right).
\label{fig:sim2}
}
\end{figure}

For the simulated dataset, we applied the proposed GWR using $\gamma$-divergence (denoted by DGWR) as well as the standard (non-robust) GWR and robust GWR using an iteratively re-weighting algorithm(denoted by RoGWR) available from R package ``GWmodel" \citep{gollini2015gwmodel}. 
The tuning parameters in DGWR are selected among 
\begin{equation}\label{candidate}
\begin{split}
\gamma\in \{0, 0.01, 0.03, 0.05 ,0.1,0.15,\ldots,0.5\}\\
b\in \{b^{\ast}/10, 2b^{\ast}/10,\ldots, 9b^{\ast}/10, b^{\ast} \},
\end{split}
\end{equation}
where $b^{\ast}$ is the sample median of $\{\|s_i-s_j\|\}_{i,j=1,\ldots,n}$.
Note that $\gamma=0$ corresponds to the standard (model-based) GWR.
We selected the optimal bandwidth for GWR via cross validation, and the same bandwidth is used for RoGWR.
These methods are performed using R package ``GWmodel" \citep{lu2014gwmodel}.
The estimation performance is evaluated based on the mean squared error (MSE) given by 
$$
{\rm MSE}=\frac1{np}\sum_{i=1}^n\sum_{k=0}^{p-1}\left\{\widehat{\beta}_k(s_i)-\beta_k(s_i)\right\}^2,
$$ 
where $p=3$ and $\widehat{\beta}_k(s_i)$ is the estimated value of $\beta_k(s_i)$. 
In Figures \ref{fig:sim1} and \ref{fig:sim2}, we show boxplots of MSE values based on 500 replications under two scenarios of outlier generation structures. 
It is observed that once outliers are included in the dataset, both RoGWR and DGWR give more stable estimations than the standard GWR under both scenarios of contamination. 
Comparing RoGWR and DGWR, they provide pretty similar performance under contamination, but it can be seen that RoGWR may break down under some scenarios such as $\omega=0.15$ under scenario (II).
Comparing the results under two settings of $\phi$, we can see that the MSE values under $\phi=0.8$ (considerable spatial collinearity) tend to be larger than those under $\phi=0.4$, but the relative performance among the three methods is almost the same. 
This means that all the models including DGWR are not robust against spatial collinearity, but DGWR can improve the performance of GWR under existence of outliers, regardless the amount of spatial collinearity.   
In Table \ref{tab:sim}, the average values of selected bandwidth and robustness parameter ($\gamma$) are presented.
From the results, we can observe that the selected values of $\gamma$ increase according to the outlier ratios, which shows that the proposed method has adaptive robustness property; the method exhibits strong robustness with non-negative $\gamma$ under the existence of outliers while it provides efficient estimation with zero value of $\gamma$ under no outliers.   
It is also observed that the difference of the bandwidth parameter selected by GWR and DGWR gets more prominent as the outlier ratio increases, and the use of the unnecessarily large value of bandwidth would lead to over-smoothed estimates of the spatially varying coefficients.     
We note that the difference in the selected bandwidth under $\omega=0$ is that DGWR estimates the spatially-varying variance parameter that the standard GWR does not estimate, resulting in the superior performance of DGWR to the other methods.

\begin{table}[!htb]
\caption{The number of locations where the condition number is greater than 30 (high spatial collinearity) among 500 locations. 
\label{tab:NC}
}
\begin{center}
\begin{tabular}{cccccccccccc}
\hline
&& \multicolumn{9}{c}{$h$ (bandwidth)}\\
 && 0.1 & 0.15 & 0.2 & 0.25 & 0.3 & 0.35 & 0.4 & 0.45 & 0.5 \\
\hline
$\phi=0.4$ && 109 & 41 & 12 & 1 & 0 & 0 & 0 & 0 & 0 \\
$\phi=0.8$ && 141 & 94 & 80 & 81 & 78 & 73 & 69 & 52 & 33 \\
\hline
\end{tabular}
\end{center}
\end{table}

\begin{table}[!htb]
\caption{Average values of selected bandwidth and $\gamma$ based on 500 replications.
\label{tab:sim}
}
\begin{center}
\begin{tabular}{cccccccc}
\hline
&&& \multicolumn{4}{c}{$\omega$ (outlier ratio)}\\
Scenario & $\phi$ &  & 0 & 0.05 & 0.1 & 0.15 \\
 \hline
 &  & bandwidth (GWR) & 0.125 & 0.245 & 0.331 & 0.434 \\
1 & $\phi=0.4$ & bandwidth (DGWR) & 0.156 & 0.171 & 0.175 & 0.183 \\
 &  & Gamma (DGWR) & 0.000 & 0.141 & 0.208 & 0.269 \\
 \hline
 &  & bandwidth (GWR) & 0.129 & 0.253 & 0.384 & 0.490 \\
1 & $\phi=0.8$ & bandwidth (DGWR) & 0.161 & 0.177 & 0.183 & 0.193 \\
 &  & Gamma (DGWR) & 0.000 & 0.139 & 0.204 & 0.265 \\
 \hline
 &  & bandwidth (GWR) & 0.125 & 0.228 & 0.313 & 0.378 \\
2 & $\phi=0.4$ & bandwidth (DGWR) & 0.156 & 0.166 & 0.172 & 0.178 \\
 &  & Gamma (DGWR) & 0.000 & 0.199 & 0.263 & 0.319 \\
 \hline
 &  & bandwidth (GWR) & 0.129 & 0.241 & 0.316 & 0.405 \\
2 & $\phi=0.8$ & bandwidth (DGWR) & 0.161 & 0.171 & 0.176 & 0.180 \\
 &  & Gamma (DGWR) & 0.000 & 0.190 & 0.247 & 0.301 \\
\hline
\end{tabular}
\end{center}
\end{table}

\section{Application}\label{sec:app}

\begin{figure}[!htb]
\centering
\includegraphics[width=15cm,clip]{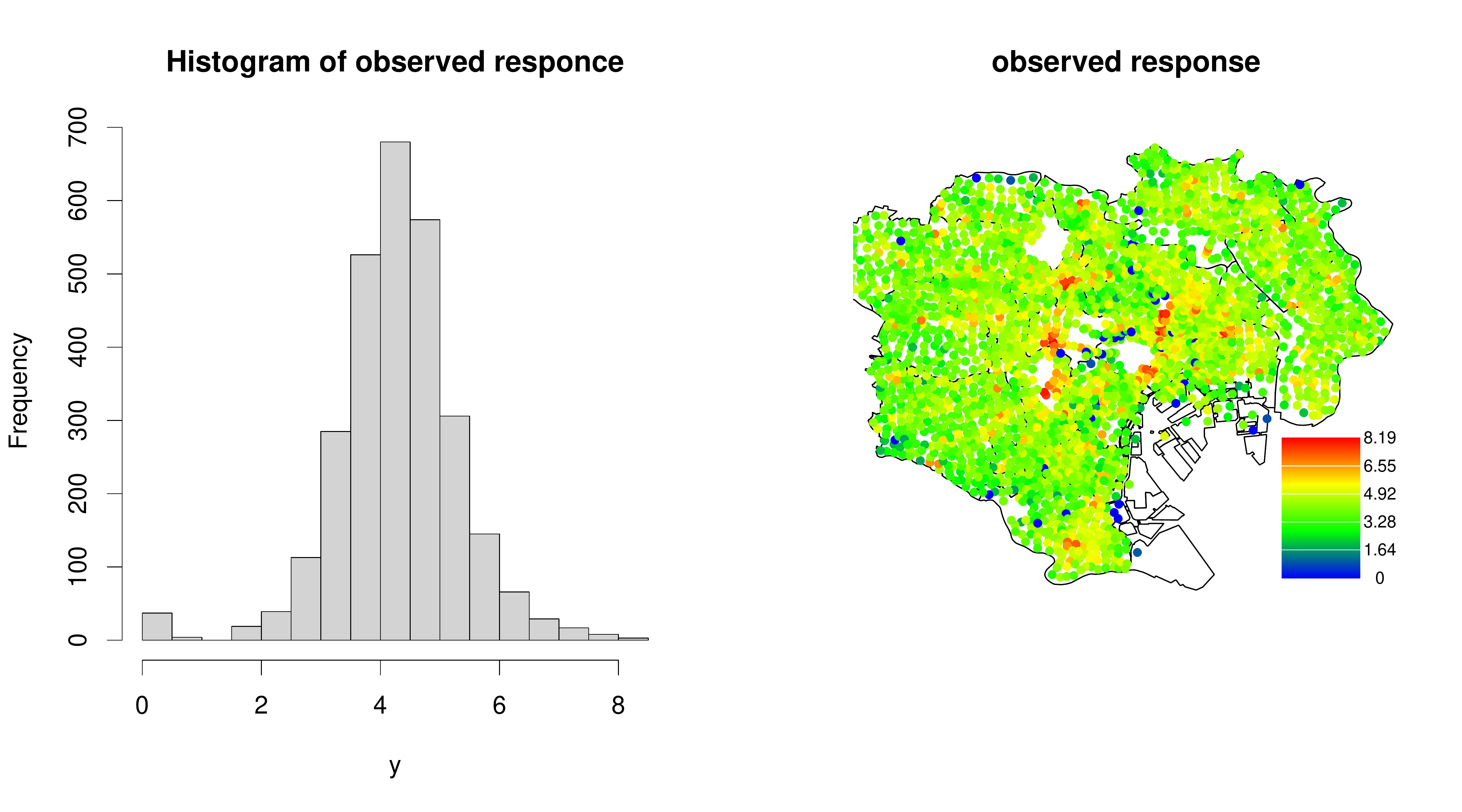}
\caption{Histogram (left) and spatial distribution (right) of the response variable.  
\label{fig:crime-obs}
}
\end{figure}

Here we apply the proposed methods to a dataset of the number of police-recorded crimes in the 23 special words of Tokyo, provided by the University of Tsukuba and publicly available online (``GIS database of the number of police-recorded crime at O-aza, chome in Tokyo, 2009-2017'', available at \url{https://commons.sk.tsukuba.ac.jp/data_en}). 
This study focuses on the number of non-burglary crimes in $n=2,855$ minor municipal district in the target area in 2015.
For auxiliary information in each district, we adopted area (km$^2$), entire population density (PD), day-time population density (DPD), the density of foreign people (FD), percentage of single-person households (SH), and average year of living (AYL).
Let $y_i^{\ast}$ be the observed count of violent crimes, $s_i$ be a two-dimensional vector of longitude and litigate of the center, $a_i$ be area (km$^2$) and $x_i$ be the vector of standardized auxiliary information in the $i$th district.
We used $y_i\equiv \log(1+y_i^{\ast}/a_i)$ to be the response variable in this analysis, which can be interpreted as logarithm values of the number of crimes per unit km$^2$. 
The histogram and spatial distribution of $y_i$ are presented in Figure \ref{fig:crime-obs}, from which we can observe that most of the observations are appeared to be normally distributed. At the same time, there are some outliers with too small or too large values of $y_i$.
It is also observed from the spatial distribution in Figure \ref{fig:crime-obs}, such extreme observations are not spatially clustered but are scattered locally.

\begin{figure}[!htb]
\centering
\includegraphics[width=15cm,clip]{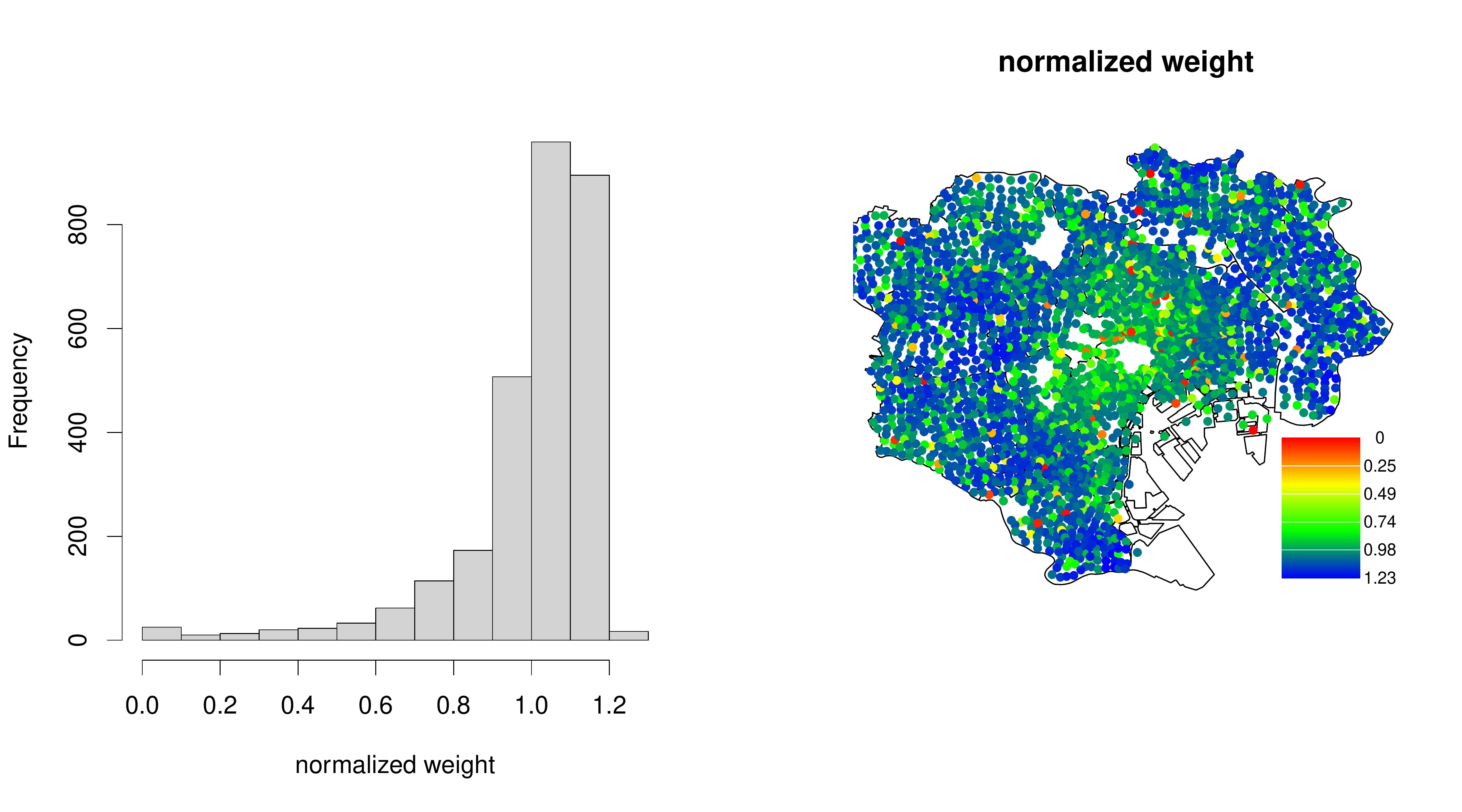}
\caption{Histogram (left) and spatial distribution (right) of the normalized weight for local outlier detection.  
\label{fig:crime-w}
}
\end{figure}

\begin{figure}[!htb]
\centering
\includegraphics[width=11cm,clip]{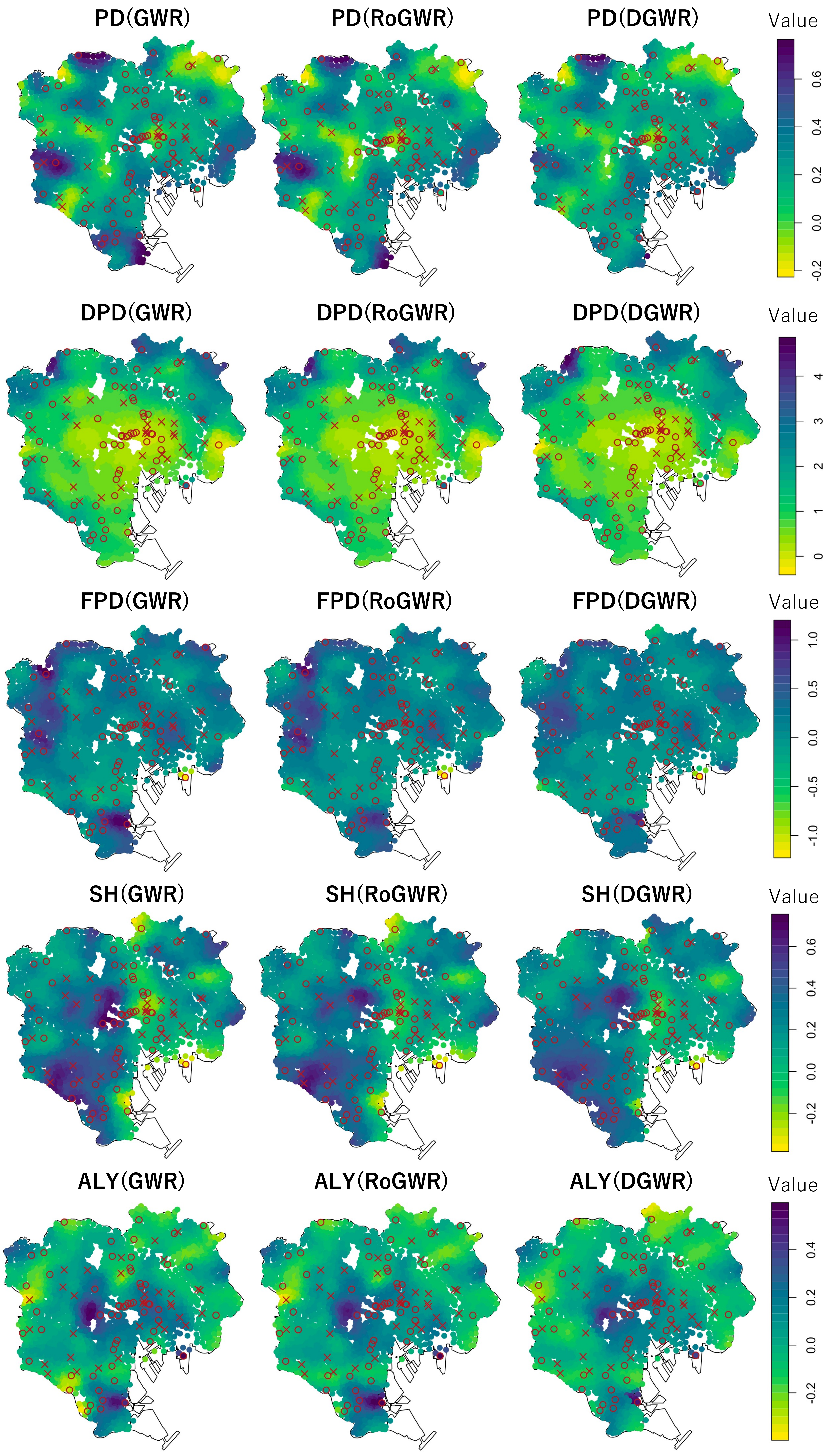}
\caption{Spatial distribution of the estimated regression coefficients based on the standard GWR, robust GWR (RoGWR), and the proposed GWR with robust divergence (DGWR). The circle and cross correspond to locations having potential outliers with large and small values, respectively, detected by DGWR.  
\label{fig:crime}
}
\end{figure}

\begin{figure}[!htb]
\centering
\includegraphics[width=15cm,clip]{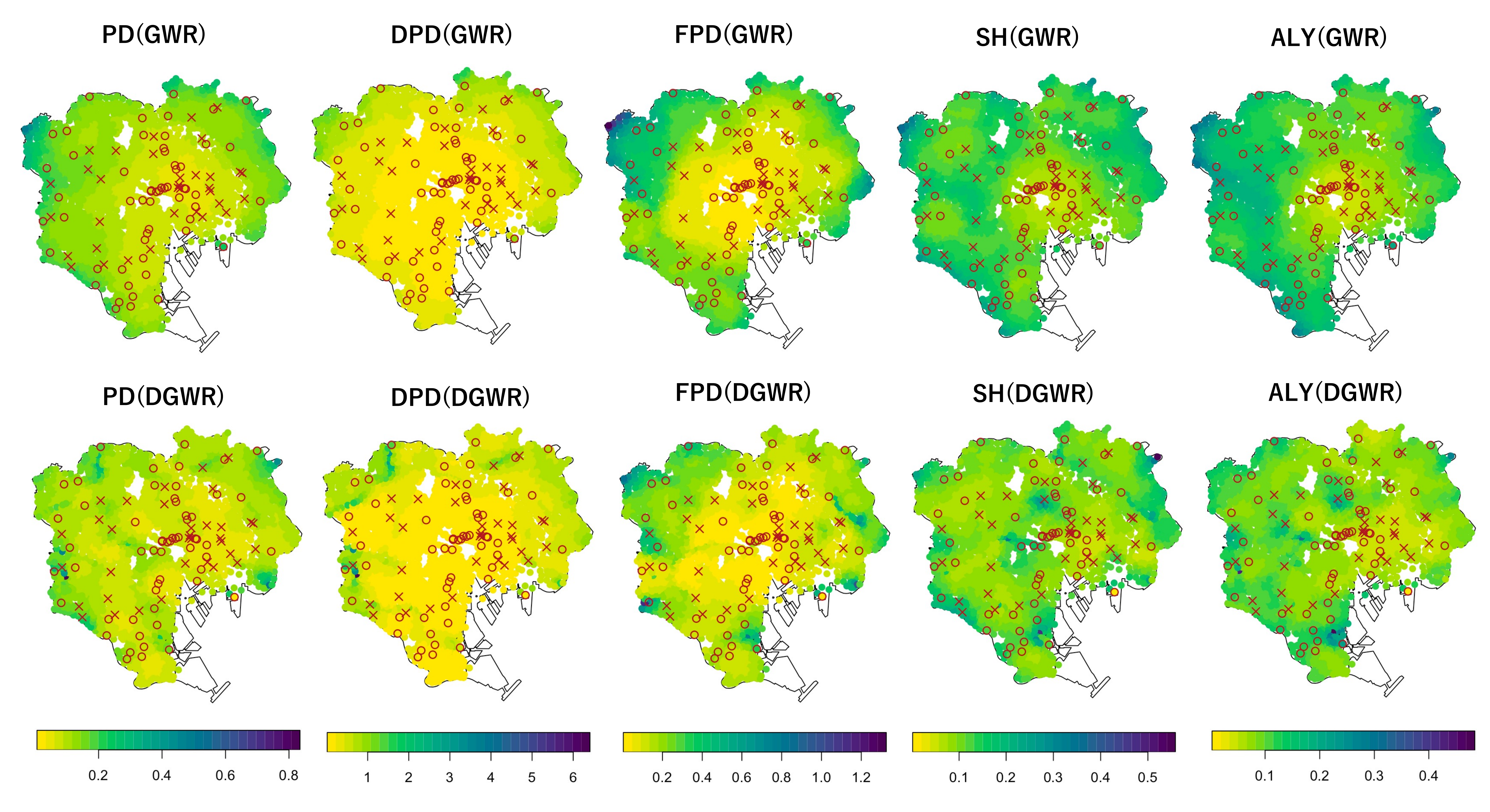}
\caption{Spatial distribution of the estimated standard errors of local regression coefficients based on GWR and DGWR. 
The circle and cross correspond to locations having potential outliers with large and small values, respectively, detected by DGWR.  
\label{fig:crime-sd}
}
\end{figure}

Using the five covariates described above, we apply the three methods adopted in the simulation study in Section \ref{sec:sim}.
We selected the optimal values of $b$ and $\gamma$ among (\ref{candidate}) in the proposed DGWR, and obtained $\gamma_{\rm opt}=0.25$ and $b_{\rm opt}=0.217$.
We also adopted different values of $b^{\ast}$ to select $\gamma$, but the optimal value of $\gamma$ did not change.
Since the selected $\gamma_{\rm opt}$ is positive, the dataset may contain outliers, which is consistent with the results given in Figure \ref{fig:crime-obs}.
The selected bandwidth for GWR and RoGWR was 0.178, which is slightly smaller than that of DGWR. 
In Figure \ref{fig:crime-w}, we provide the histogram and spatial distribution of the normalized weight (\ref{weight}) for local outlier detection. 
Recall that most normalized weights are around 1, corresponding to non-outliers, but a substantial number of samples have small normalized weights, which are likely to be local outliers. 
In this analysis, we systematically regarded samples whose normalized weights are smaller than 0.5 as local outliers. 
In the figure \ref{fig:crime}, we presented spatial distributions of estimated spatially-varying regression coefficients for the five covariates, together with the locations of detected local outliers.
We can observe that GWR and RoGWR give relatively similar results because they used the same bandwidth. In contrast, the proposed DGWR provides almost the same but slightly different results, especially around the locations with local outliers, for example, the south part of FPO and ALY and the southwest part of SH.
We also computed standard errors of the estimates, using the formula in Section \ref{sec:SE} for the proposed DGWR method and R package ``spgwr" \citep{spgwr} for GWR. 
The results are shown in Figure \ref{fig:crime-sd}, from which apparent differences between standard errors in GWR and those in DGWR can be confirmed. 
It can be seen that the standard errors in GWR tend to be large around locations in which samples are identified to be outliers, which would indicate that the standard GWR fails to carry out reasonable uncertainty quantification due to the influence of outliers.

\section{Conclusion and Discussion}\label{sec:disc}

We employed a technique of $\gamma$-divergence to develop a robust version of GWR. 
A novel feature of the proposed method is its adaptive property; the two tuning parameters that control robustness and spatial smoothness are automatically determined through reasonable criteria. 
The simulation and empirical studies showed the superiority of the proposed method to the existing methods.

Although the proposed method is entirely focused on the spatially varying linear regression, the proposed framework for robust fitting of the spatially varying models can be generalized to more general regression models such as generalized linear models. 
Let $f(f_i|x_i;\theta_i)$ be a regression model in the $i$th location. 
Then, the objective function is given by
\begin{equation}\label{general}
D_i(\theta)=\frac1{\gamma}\log\left(\sum_{j=1}^n w_{ij}(b) f(y_j|x_j;\theta)^{\gamma}\right) +\frac1{1+\gamma}\log\left\{\sum_{j=1}^n w_{ij}(b) \int f(t|x_j;\theta)^{1+\gamma}dt\right\}, 
\end{equation}
and the estimator of $\th_i$ can be defined as the maximizer of the above objective function. 
When $y_i\sim N(x_i^{\top}\beta_i, \sigma_i^2)$, the general objective function reduces to (\ref{gam-div}) for the proposed DGWR.
The bandwidth parameter can be selected by minimizing the following criterion: 
$$
{\rm RCV}(b; \gamma) = \frac1{\gamma}\log\left(\sum_{i=1}^n  f(y_i|x_i;\thh_{i(-i)})^{\gamma}\right) 
+\frac1{1+\gamma}\log\left\{\sum_{i=1}^n \int f(y_i|x_i;\thh_{i(-i)})^{1+\gamma}dt\right\},
$$
where $\thh_{i(-i)}$ is the robust estimator of $\th_i$ without using the sample in the $i$th location, and it again reduces to (\ref{RCV}) when $y_i\sim N(x_i^{\top}\beta_i, \sigma_i^2)$. 
Moreover, the general form of the Hyvarinen score (\ref{H-score}) to select $\gamma$ is given by 
\begin{align*}
H(\gamma)
=\sum_{i=1}^n \delta(\thh_i)\bigg[2f'(y_i;\thh_{i})^2f(y_i;\thh_{i})^{\gamma-2}\left\{2(\gamma-1)+f(y_i;\thh_{i})^{\gamma}\right\}+2f(y_i;\thh_{i})^{\gamma-1}f''(y_i;\thh_{i})\bigg],
\end{align*}
where $f'(y_i;\th)=\partial f(y_i;\theta)/\partial y_i$, $f''(y_i;\th)=\partial^2 f(y_i;\theta)/\partial y_i^2$ and $\delta(\th)=\big\{\int f(t;\theta)^{1+\gamma}dt\big\}^{-\gamma/(1+\gamma)}$.
When $y_i\sim N(x_i^{\top}\beta_i, \sigma_i^2)$, the above general expression reduces to (\ref{H-score}).
Hence, we can also develop a robust version of other types of geographically weighted regression such as geographically weighted negative binomial regression \citep{da2014geographically}.
However, the main difficulty in applying the general idea is the integral appeared in the objective function (\ref{general}).
Although the integral can be analytically computed under normal distribution, it may need to rely on numerical approximation under general distributions, leading to computationally burdensome to maximize the objective function.

Finally, a potential limitation of the proposed method is its scalability under large sample sizes. 
Since the proposed method requires estimation of location-wise parameters via iterative methods, it would be computationally demanding under massive spatial data. 
Exploration of some scalable versions of DGWR, as done for the standard GWR \citep[e.g.][]{li2019fast,murakami2020scalable}, is an important and interesting issue for future studies.

\section*{Acknowledgements}
This work was supported by the Japan Society for the Promotion of Science (KAKENHI) Grant Numbers 18H03628, 20H00080, and 21H00699.

\vspace{5mm}
\bibliography{ref}
\bibliographystyle{chicago}

\end{document}